# What Does It Mean to Use AI Critically? Unpacking Critical AI Literacy Through Students' Evaluation of AI-Generated Content

Hyejeong Lee
The University of Texas at San Antonio
hyejeong.lee@utsa.edu

Wonjin Yu
The University of Alabama
wyu21@ua.edu

**Abstract**

Although critical AI literacy has emerged as an important educational goal, the construct remains conceptually broad and insufficiently specified for guiding students' day-to-day interactions with AI. This study examines how students critically evaluated AI-generated content. Drawing on an analysis of students' chatbot interactions and written reflections, we first identified seven stages of AI-supported academic task completion and five functional roles assumed by AI. More importantly, we identified eight evaluative lenses that students used to assess AI-generated responses (accuracy, completeness, task alignment/relevance, personalization, practicality, creativity, ethics, and bias). The findings suggest that critical AI literacy extends beyond verifying the factual accuracy of AI-generated information. Rather, critically engaging with AI involves situated and multidimensional judgment about the epistemic quality, contextual fit, practical feasibility, creative value, and ethical implications of AI-generated content. The framework offers both a conceptual contribution to the emerging literature on critical AI literacy and a practical tool for educators seeking to help learners move from passive acceptance of AI toward deliberate, context-sensitive, and responsible human judgment.

# 1. Introduction

Generative artificial intelligence (GenAI) has rapidly become integrated into students' academic work (Giannakos et al., 2025; Holmes & Miao, 2023; Lee et al., 2025; Su, J., & Yang, 2023; Xiao et al., 2019). Unlike many earlier educational technologies, conversational GenAI enables iterative interaction in which students can question, inspect, and respond to AI-generated suggestions (Hwang & Chang 2023; Winkler & Söllner, 2018). This emerging form of human-AI collaboration creates new opportunities for learning while raising an important question: what does it mean for students to use AI critically?

Critical engagement with GenAI requires more than the ability to generate effective prompts or incorporate AI-generated content (Xiao et al., 2026). AI responses may be fluent and plausible while being inaccurate, incomplete, irrelevant to a particular task, insufficiently tailored to a learner's context, or problematic from pedagogical perspectives (Xiao et al., 2026). Students therefore need to exercise judgment about when, why, and how AI-generated suggestions should be used. Although AI literacy scholarship increasingly recognizes evaluation and ethical judgment as important competencies (Biagini, 2025; Ng et al., 2021), less is known about how critical evaluation is enacted during students' actual academic interactions with AI.

This gap is particularly important because students' observable decisions may not adequately capture the quality of their critical engagement. Understanding critical engagement requires attention not only to students' decisions, but also to the reasoning underlying those decisions. To examine this process, we developed an AI-supported learning tool that embeds an explicit evaluation step into students' interactions with a conversational chatbot. Drawing on students' interaction logs and written decision rationales, this study examines how students engage with GenAI during academic task completion, how they make decisions about AI-generated suggestions, and the evaluative criteria underlying those decisions. Accordingly, this study asks: *How do students exercise evaluative agency when collaborating with generative AI during academic task completion?*

Specifically, we investigated three research questions.

RQ1. How do students engage with generative AI during academic task completion?

RQ2. How do students make decisions about AI-generated suggestions, and how does their evaluative engagement differ between students who accept and adapt those suggestions?

RQ3. What evaluative criteria do students use when critically assessing AI-generated suggestions?

By examining interaction patterns and evaluative reasoning, this study contributes empirical insight into how critical AI literacy is enacted during AI-supported academic work.

# 2. Literature Review

## 2.1. From Passive AI Use to Agency in Human-AI Collaboration

AI has increasingly been positioned not simply as a technological tool but as an interactive partner in learning and academic work (Casal-Otero et al., 2023; Holmes & Tuomi, 2022; Luckin, & Holmes, 2016; Touretzky et al., 2019). Unlike conventional educational technologies that primarily support learners in accessing or producing information, conversational GenAI enables iterative exchanges in which students can ask questions, solicit suggestions, evaluate responses, and revise their work (Giannakos et al., 2025; Mittal et al., 2024). In this sense, GenAI can function as a learning partner whose contributions are continuously shaped by learners' prompts, judgments, and decisions. The distinction between AI use and meaningful human-AI collaboration is therefore particularly important in educational contexts. Rather than merely receiving an AI-generated answer, learners may actively participate in negotiating what information or suggestions to incorporate into a final product (Lee et al., 2026).

This perspective aligns with broader conceptualizations of AI literacy that emphasize the ability to interact and collaborate effectively with AI rather than merely operate AI-enabled technologies. Long and Magerko (2020), in a foundational review of AI literacy, conceptualized AI literacy as a set of competencies that enables individuals to critically evaluate AI technologies, communicate and collaborate effectively with AI, and use AI as a tool across everyday contexts. Their framework highlights that AI literacy extends beyond technical knowledge to include understanding AI capabilities and limitations, interpreting AI outputs, and making informed judgments about AI-supported activities.

Similarly, Ng et al. (2021) synthesized prior scholarship on AI literacy into four broad dimensions: understanding AI, using AI, evaluating AI, and considering ethical issues. These frameworks are particularly relevant to higher education because students increasingly encounter AI not as an isolated technological artifact but as an active participant in academic work. The critical question is therefore not simply whether students possess sufficient technical competence to prompt an AI chatbot, but whether they can strategically position AI within their own thinking and maintain agency over the resulting work.

Recent studies found that students were generally able to use GenAI for a range of academic activities, while difficulties in critical evaluation of AI-generated outputs remained (Li et al., 2025; Xiao et al., 2026). This distinction between functional use and critical use is central to the present study. Students may be highly active in their interactions with AI while still engaging with AI-generated information in relatively uncritical ways (Kulal, 2025). Accordingly, human-AI collaboration in education should be understood as a process of distributed decision making in which students determine when and how AI-generated suggestions should be incorporated into their work (Lee et al., 2026). From this perspective, the student's use of AI is not merely a behavioral outcome. It represents a context of human judgment in which the learner evaluates the usefulness and appropriateness of an AI-generated contribution.

### 2.2. Critical AI Literacy and the Evaluation of AI-Generated Outputs

Critical literacy has long positioned learners as active interpreters of information rather than passive recipients of knowledge. Rooted in critical pedagogy, this perspective emphasizes learners' capacity to question taken-for-granted assumptions, examine the social and cultural conditions underlying texts and knowledge, and develop informed judgments about their implications (Freire, 1970; Luke, 2012; Shor, 1993). As AI increasingly mediates how learners encounter, produce, and evaluate information, these critical literacy practices become particularly consequential. AI-generated outputs can appear authoritative while reflecting the data, assumptions, and design choices embedded in AI systems. Consequently, learners cannot be positioned merely as consumers of AI-generated information (Velander et al., 2024). Instead, they need to interrogate, evaluate, and, when necessary, challenge AI outputs while exercising judgment over how such outputs are used.

AI literacy has been broadly conceptualized as a set of competencies that enables individuals to understand, use, evaluate, communicate about, and collaborate with AI (Long & Magerko, 2020; Miles & Lee, 2026; Ng et al., 2021). Importantly, these conceptualizations extend beyond technical knowledge or functional proficiency to include the ability to critically examine AI technologies and their outputs (Lee et al., 2026). Critical evaluation is particularly consequential in educational contexts because AI-generated content is not inherently authoritative simply because it is fluent, coherent, or responsive. Learners must be able to interrogate AI-generated information and make informed judgments about how, when, and whether such content should be used.

Recent scholarship has begun to more explicitly conceptualize critical AI literacy as a distinct dimension of AI literacy. For example, Veldhuis et al. (2025) examined critical AI literacy in relation to learning objectives and identified three broad cognitive dimensions: analyzing, evaluating, and creating. These dimensions encompass examining how AI is integrated into everyday life and shapes identity, evaluating how AI is applied and its potential applications, and envisioning responsible ways of using AI technologies. Ng et al. (2021) also identified evaluation as one of the major dimensions of AI literacy, requiring learners to analyze, evaluate, and justify decisions about AI-generated outputs. Such conceptualizations establish critical engagement as a central component of AI literacy and emphasize learners' active judgment rather than passive consumption of AI technologies.

However, the notion of critical evaluation remains relatively broad in the existing literature. In educational settings, evaluating AI-generated content involves more than determining whether an output is factually correct. An AI-generated response may be factually accurate yet still be unsuitable for a particular learner, discipline, audience, or instructional context. Thus, critical engagement with AI requires learners to consider multiple dimensions of an AI response and determine whether, how, and to what extent it should inform their own work.

Studies have identified critical evaluation and the appropriate incorporation of AI-generated content as important dimensions of GenAI literacy while also documenting challenges in learners' ability to systematically evaluate AI outputs (Tang et al., 2025; Velander et al., 2024; Veldhuis et al., 2025). Collectively, this literature establishes critical evaluation as an important component of AI literacy and highlights the need for learners to exercise judgment when engaging with AI-generated content. Yet, the construct remains articulated as a broader competency than as a set of concrete evaluative practices that can be observed in learners' interactions with AI.

### 2.3. Research Gap and Aim

Despite growing recognition of critical evaluation as a central component of AI literacy, the empirical manifestation of critical AI literacy during actual human-AI interaction remains insufficiently understood. Although critical evaluation is increasingly recognized as an important dimension of AI literacy, less is known about how learners actually evaluate AI-generated suggestions.

This gap is consequential because students' observable decisions alone may provide an incomplete picture of critical engagement. Critical engagement may become visible when a student identifies limitations in an AI response, requests clarification, compares alternatives, evaluates the response against task-specific criteria, or modifies the output based on contextual considerations. Accordingly, there is a need for empirical research that moves beyond measuring whether or how much learners use AI to examine how learners evaluate AI-generated content and exercise judgment in response to it. In particular, greater attention is needed to the evaluative criteria through which students assess AI-generated suggestions in authentic academic contexts. Identifying such criteria can help make an otherwise broad and abstract construct of critical AI literacy more observable and empirically examinable.

Addressing this gap, the present study examines students' human-AI interactions during authentic academic task completion, with particular attention to their engagement with and decision-making about AI-generated suggestions. Specifically, the study examines the ways students position AI across different stages of academic task completion and investigates the reasoning underlying their decisions to accept or adapt AI-generated suggestions. Most importantly, the study identifies the evaluative lenses through which students assess AI-generated content, providing an empirically grounded account of how critical AI literacy is enacted through human-AI interaction. In doing so, this study seeks to move the conceptualization of critical AI literacy from a broad normative expectation toward an empirically grounded account of what learners evaluate and how they exercise judgment when doing so.

## 3. Method

### 3.1. Research Design

We employed a mixed-methods research design (Creswell, 1999; Tashakkori & Creswell, 2007) to examine how students interacted with generative AI and evaluated AI-generated suggestions during academic task completion. Quantitative data were used to examine students' AI evaluation and satisfaction across decision conditions, while qualitative data provided insight into students' evaluative reasoning and the criteria they used when deciding how to incorporate AI-generated suggestions.

### 3.2. Participants and Context

Participants were 44 students from two universities in the southern United States. The sample included 32 master's students enrolled in an instructional technology graduate course at University A and 12 undergraduate teacher candidates from various subject-area majors enrolled in a technology-related course at University B. All participants used the AI chatbot as part of a course-related final project, and participation in the research was voluntary. Students could choose whether to use the chatbot to support their project work and could decline to participate in the study or withdraw without any effect on their course grades.

The research team developed a web-based generative AI chatbot using Google AI Studio to support students' academic project work. Rather than functioning primarily as an answer-delivery system, the chatbot was designed as a project-focused thinking partner that supported iterative student-AI interaction.

At the beginning of an interaction, students selected the type of task they were working on, allowing the chatbot to tailor its responses to the project context. The chatbot generated suggestions in concise, structured segments and concluded substantive responses with a follow-up question intended to encourage students to clarify, extend, or reconsider their ideas. Importantly, the system incorporated an explicit evaluation step into the interaction. After receiving an AI-generated suggestion, students were asked to indicate whether they would Accept or Adapt the suggestion and to provide a written justification for their decision, including the criteria they considered in evaluating the suggestion.

### 3.3. Data Sources

Data were collected from students' interactions with the AI chatbot and included student-AI conversation logs, conversation histories, and students' written justifications for their decisions regarding AI-generated suggestions. Across the 44 participants, the dataset contained 229 user turns, corresponding to an average of 5.2 user turns per participant ($M = 5.2$, $Mdn = 3$, range = 1-22). Multiple exchanges occurred in 34 of the 44 conversations (77.3%), indicating that most students engaged in iterative rather than single-turn interactions with the chatbot. Quantitative data included students' Information Evaluation (IE) scores, measured using the previously developed and validated scale (e.g., critically evaluating and substantiating relevance, credibility, and logical strength in GenAI responses) by Li et al. (2025), which assesses critical interaction with AI. Students also reported their satisfaction with the chatbot's outputs following the interaction.

### 3.4. Data Analyses

Qualitative data were analyzed using inductive thematic analysis (Braun & Clarke, 2006; Clarke & Braun, 2017) approach. The analysis focused on students' written rationales for accepting or adapting AI-generated suggestions. Rather than applying a predetermined coding framework, codes were developed inductively from students' responses to capture the evaluative criteria reflected in their reasoning. Through an iterative process of coding, comparison, and refinement, conceptually related codes were grouped into broader themes representing the lenses through which students evaluated AI-generated suggestions. These themes were subsequently used to characterize dimensions of critical AI literacy manifested in students' evaluative reasoning. Quantitative data were analyzed using independent t-tests in their IE scores and satisfaction ratings.

## 4. Results

### 4.1. Patterns of AI Use during Academic Task Completion

First, we examined how students engage with generative AI during academic task completion. Our analysis revealed seven distinct stages at which students engaged with the chatbot during academic task completion: explanation, planning, brainstorming, draft generation, revision, evaluation, and reflection (Table 1). These stages should not be interpreted as a fixed or linear sequence through which all students progressed. Rather, students selectively engaged with the AI chatbot at different points in the task completion process, with some students using the chatbot across multiple stages and others drawing on it for only a subset of activities. Thus, the seven stages represent the range of academic activities for which students mobilized AI support rather than a universal progression followed by every student.

Across these stages, the AI chatbot served five primary functional roles: tutor, idea generator, decision supporter, editor, and evaluator. These two dimensions capture complementary aspects of students' AI use. The stages characterize *when and for what type of academic activity* students engaged with AI, whereas the functional roles characterize *what the AI was expected to do* within those activities.

**Table 1**
*Overview of Each Stage, AI's Role, and Student Prompt*

| Stage | AI Functional Role | Explanation | Example Student Prompt |
|---|---|---|---|
| Explanation | Tutor | Seeking explanations or clarification to understand concepts, content, or task requirements | "What are the most common performance gaps for and how should I address them in training?" |
| Planning | Decision Supporter | Establishing the overall direction, structure, or approach for completing the task | "I am creating an instructional design plan with three modules..." |
| Brainstorming | Idea Generator | Generating new ideas, possibilities, examples, or alternative approaches | "Could you give me some activity suggestions for my lesson?"<br>"I need ideas for how to make..." |
| Draft generation | Idea Generator /Editor | Generating initial content or a preliminary version of the task product | "Create a survey to determine how well students liked their training modules." |
| Revision | Editor | Revising, refining, reorganizing, or improving existing content or work | "I would like assistance improving a lesson activity..." |
| Evaluation | Evaluator | Seeking feedback or assessment of the quality, appropriateness, or effectiveness of an idea or product | "Is this a good objective including Bloom's taxonomy?" |
| Reflection | Decision Supporter/ Evaluator | Reviewing AI suggestions and reflecting on whether, how, or why they should be incorporated into one's work | "I chose to adapt the AI suggestions because…" |

#### *4.1.1. Explanation and Planning: AI as Tutor and Decision Supporter*

Students engaged with the AI agent during the early phases of task completion to develop conceptual understanding, clarify task requirements, and determine possible directions for their work. In these interactions, the AI primarily functioned as a tutor or decision supporter. As a tutor, the chatbot was used to explain concepts, clarify unfamiliar content, or help students understand what a task required. As a decision supporter, it helped students consider possible directions, structures, or approaches before committing to a particular course of action.

For example, one student used the chatbot to inquire about "the most common performance gaps" and how these gaps could be addressed through training. Rather than requesting a finished product, the student used the chatbot to establish an initial understanding of the problem space. Similarly, students described their plans to the chatbot and sought input on how to structure their instructional design work. These interactions suggest that AI was sometimes positioned as a resource for orienting students to a task and expanding their consideration of possible approaches, rather than simply producing an answer.

#### *4.1.2. Brainstorming and Draft Generation: AI as Idea Generator*

As students moved from understanding and planning toward developing their own products, AI was frequently used for brainstorming and draft generation. In this stage, the chatbot primarily functioned as an idea generator, providing possible topics, activities, examples, alternatives, or preliminary content that students could consider. Students' prompts included requests such as, "Could you give me some activity suggestions for my lesson?" and "I need ideas for how to make videos..." Other students asked the chatbot to generate an initial product, such as a survey for evaluating a training module. These interactions indicate that students used AI not only to obtain information but also to expand the range of possibilities available to them during the ideation process.

At the same time, draft generation was not necessarily synonymous with accepting AI-generated content as a final product. Students frequently treated generated content as a starting point that required subsequent modification, evaluation, or contextualization. Thus, the generation of content often initiated additional stages of human judgment rather than concluding the task.

#### *4.1.3. Revision: AI as Editor*

Students also engaged with AI after developing an initial version of their work. In these interactions, the chatbot primarily functioned as an editor, assisting students in revising, refining, reorganizing, or improving existing content. Rather than asking the chatbot to independently produce an entire task product, students presented existing ideas or materials and requested assistance in improving specific components.

For example, one student asked the chatbot for assistance with improving a lesson activity. This pattern suggests that students used AI as a form of iterative writing and design support, in which the student retained an existing product or idea while using AI to identify possible improvements. The distinction between draft generation and revision is therefore important. Whereas draft generation involved using AI to produce preliminary content, revision involved using AI to transform or refine content that students had already developed.

#### *4.1.4. Evaluation and Reflection: AI as Evaluator and Decision Supporter*

Later-stage interactions involved students using AI to assess ideas, review existing work, and consider whether AI-generated suggestions should be incorporated into their final products. In these interactions, the AI agent primarily functioned as an evaluator and decision supporter.

Students asked questions such as, "Is this a good objective including Bloom's taxonomy?" to obtain feedback on the quality or appropriateness of their work. Other interactions involved students explicitly reflecting on AI-generated suggestions and deciding whether to accept or reject them. Reflection therefore represented a particularly important point of human judgment: students were not merely receiving AI output but considering how that output should influence their own decisions.

These findings demonstrate that students' use of AI extended across multiple activities within academic task completion, from understanding and planning to generation, revision, evaluation, and reflection. The functional roles assumed by the chatbot similarly varied according to students' immediate needs. Rather than using AI exclusively as an answer-generating mechanism, students positioned the chatbot as a tutor, source of ideas, decision-support resource, editor, and evaluator at different points in their work. This variation highlights the functional flexibility of AI in academic task completion and provides a basis for examining the next question: *How did students determine whether AI-generated suggestions were worth using?* We address this question through the evaluation lenses.

### 4.2. Students' Decisions about AI-Generated Suggestions and Group Differences

Next, we compared students who accepted AI-generated suggestions with those who adapted them when doing their assignments. We examined group differences in IE scores and satisfaction with AI responses. The descriptive and inferential statistics are presented in Tables 2, 3, and Figure 1. Students who accepted AI-generated suggestions reported significantly higher satisfaction with the AI responses ($M = 4.63$, $SD = 0.62$) than students who adapted the suggestions ($M = 4.14$, $SD = 0.76$), $t(42) = 2.17$, $p = .036$, $d = 0.68$. This represents a moderate-to-large effect, suggesting that students who ultimately accepted AI-generated suggestions tended to perceive those responses as more satisfactory. In contrast, there was no statistically significant difference in information evaluation scores between students who accepted ($M = 1.44$, $SD = 1.03$) and adapted ($M = 1.61$, $SD = 1.26$) AI suggestions, $t(42) = -0.46$, $p = .649$, $d = -0.14$. Thus, students who modified AI-generated suggestions did not demonstrate significantly higher information evaluation scores than those who accepted the suggestions.

**Table 2**

*Group Descriptives*

| Construct | Group | Mean | SD | SE |
|---|---|---|---|---|
| Satisfaction | Accept | 4.63 | 0.62 | 0.155 |
| | Adapt | 4.14 | 0.76 | 0.143 |
| IE Score | Accept | 1.44 | 1.03 | 0.258 |
| | Adapt | 1.61 | 1.26 | 0.238 |

**Table 3**
*Independent T-Test*

| Construct | t | df | *p* | d |
|---|---|---|---|---|
| Satisfaction | 2.167 | 42 | .036 | 0.679 |
| IE Score | -0.458 | 42 | .649 | -0.144 |

**Figure 1**
*Satisfaction and IE Score Plots*

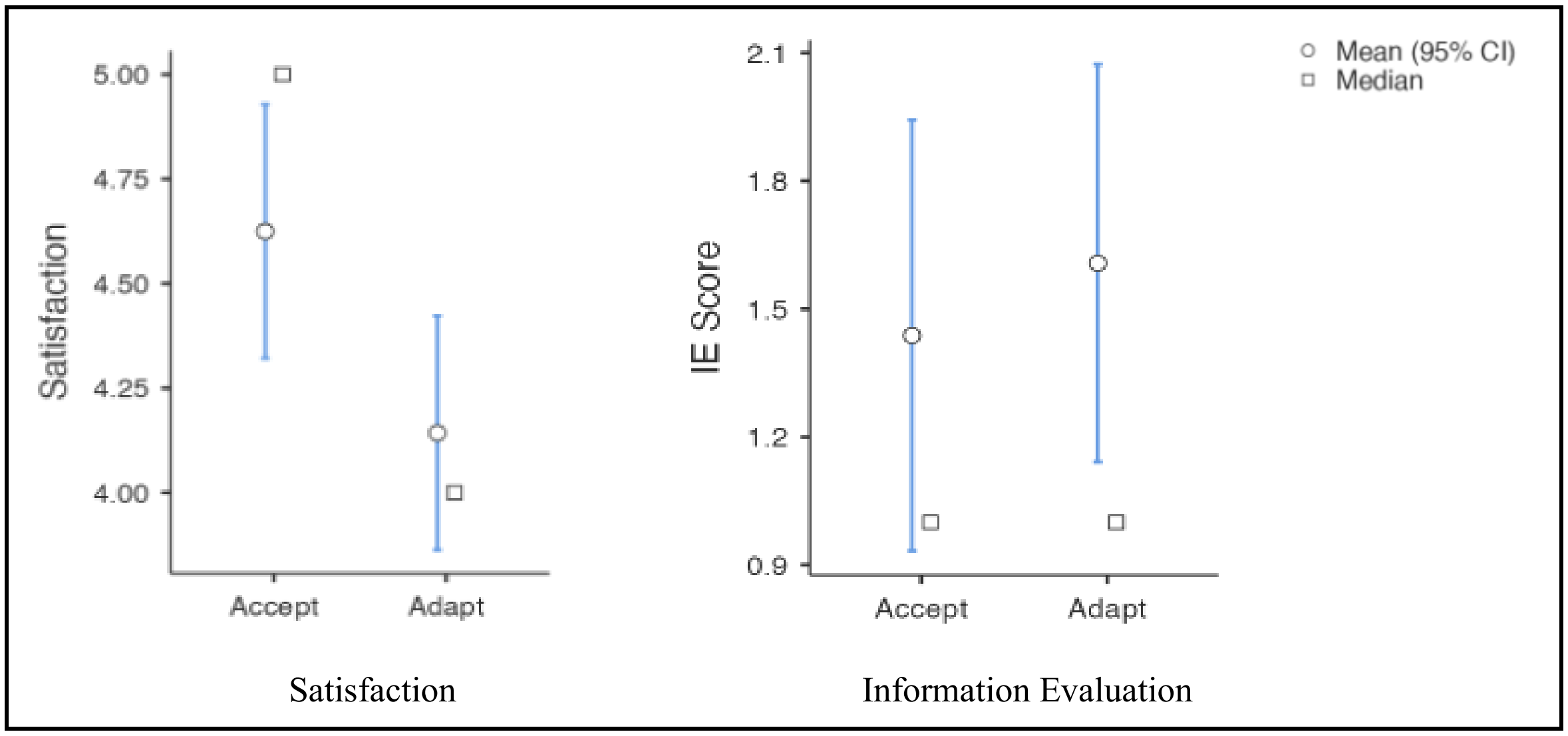

Then, we examined students' decision justification. Students who accepted AI-generated suggestions often described their decisions as grounded in the perceived usefulness and appropriateness of the responses. One student explained, *"I found the responses to be appropriate and helpful."* Another noted, *"I did use AI's suggestions it gave me because they helped me see areas for improvement and potential situations that might come up, and it gave me solutions."* These responses indicate that acceptance could reflect an evaluative judgment that an AI-generated suggestion sufficiently met the student's needs rather than simple, unquestioned compliance with the chatbot.

Similarly, students who adapted AI-generated suggestions frequently reflected an attempt to contextualize or refine an otherwise useful suggestion. One student described the AI recommendations as potentially beneficial while acknowledging the need for further testing: *"The recommendations could enhance my presentation, but I won't know until I complete test trials."* This response illustrates an evaluative process in which the student provisionally considered the AI suggestion while recognizing that its usefulness could only be determined through practical implementation. Other students explicitly articulated the criteria underlying their adaptations. For example, one student stated, *"I used the general principles of relevance* *and accuracy to evaluate the AI tools response. I used these to evaluate how the response given related to the prompt and course content given to the chatbot."* Here, adaptation was not simply a behavioral response to dissatisfaction with the AI output. Rather, the student evaluated the response against explicit criteria (e.g., relevance and accuracy) and used those judgments to determine how the response should be incorporated. The student retained the generative value of the AI suggestion while exercising independent judgment about its alignment with the specific learning context and assessment purpose.

These results indicate that acceptance and adaptation represent outcomes of students' decision making rather than direct indicators of the criticality of that decision making. Students who accepted AI suggestions sometimes articulated reasons for doing so, including perceived appropriateness and usefulness, while students who adapted suggestions often described retaining AI-generated ideas while modifying them to satisfy contextual, practical, or task-specific requirements. Consequently, the qualitative reasoning underlying students' decisions provided richer evidence of evaluative engagement than the decision category itself.

These findings motivated a closer examination of what criteria students actually used to make these judgments. Analysis of students' chatbot interactions and written reflections revealed eight recurring evaluative lenses through which students assessed AI-generated suggestions. These lenses are presented in Section 4.3.

### 4.3. Evaluation Lenses for Critical AI Literacy

Finally, to examine how students exercised judgment when interacting with AI, we analyzed the rationales underlying their decisions to accept or modify AI-generated responses. Across the chatbot interactions and students' written reflections, we identified eight recurring evaluation lenses through which students assessed the usefulness, appropriateness, and implications of AI-generated content: accuracy, completeness, task alignment/relevance, personalization, practicality, creativity, bias/fairness, and ethics/responsibility (Table 4, Figure 2).

These lenses indicate that students did not uniformly treat AI-generated responses as authoritative or inherently appropriate. Rather, students brought their own evaluative criteria to the interaction and assessed AI output across multiple dimensions. Importantly, students' evaluations extended beyond conventional concerns about factual accuracy. They considered whether AI responses were sufficiently complete, aligned with the purposes and requirements of their assignments, appropriate for their intended audiences and contexts, practically feasible, sufficiently original, inclusive and unbiased, and ethically responsible. The eight lenses can be organized into four broader dimensions: epistemic (accuracy, completeness), contextual (task alignment/relevance, personalization), practical/design (practicality, creativity), and ethical/social (bias/fairness, ethics/responsibility). The following sections describe each lens and illustrate how students articulated these judgments.

**Table 4**

*Evaluation Lens for Critical AI Literacy*

| Dimension | Evaluation Lens | Evaluation Question | Sample Student Quote |
|---|---|---|---|
| Epistemic criteria | Accuracy | Is the information accurate and factually reliable? | "I was looking for accuracy to the topic. Honestly, I was really open to all suggestions with the intention to fact check and consider the logistics of certain implementations afterwards." "It explained the concept well." |
| | Completeness | Is the response sufficiently detailed, specific, and comprehensive? | "Need more examples." "AI's suggestions were detailed" "I'd like to check whether it would provide me with enough information." |
| Contextual criteria | Task Alignment/ Relevance | Is the response relevant to the assignment and aligned with its goals or requirements? | "I need relatability for my assignment" "I wanted to check the relevance." "This isn't aligned with my assignment objective." |
| | Personalization | Does the response fit the intended audience, learner, or specific context? | "I need it to match my audience." "I evaluated the AI suggestions based on my target audience." |
| Practical/ Design criteria | Practicality | Is the AI's suggestion feasible and applicable in practice? | "Not realistic." "This wouldn't work." "The activity wasn't appropriate." |
| | Creativity | Does the response offer original, diverse, or non-generic ideas? | "I want something different." "Too generic." |
| Ethical criteria | Ethics/ Responsibility | Is the use or application of the AI response ethically and responsibly appropriate? | "I examined the responses whether it would cross an ethical line to investigate the subject on my own." "I really appreciated the response, including a paragraph on the ethical and legal aspects of the subject. I think this is critical when it comes to potential abuse." |
| | Bias/Fairness | Is the response unbiased, fair, and inclusive of diverse perspectives and groups? | "Is it neutral or biased? "What are your thoughts for the most inclusive experience?" "This could be a disadvantage…To ensure an inclusive experience…" |

**Figure 2**

*Critical AI Literacy Framework*

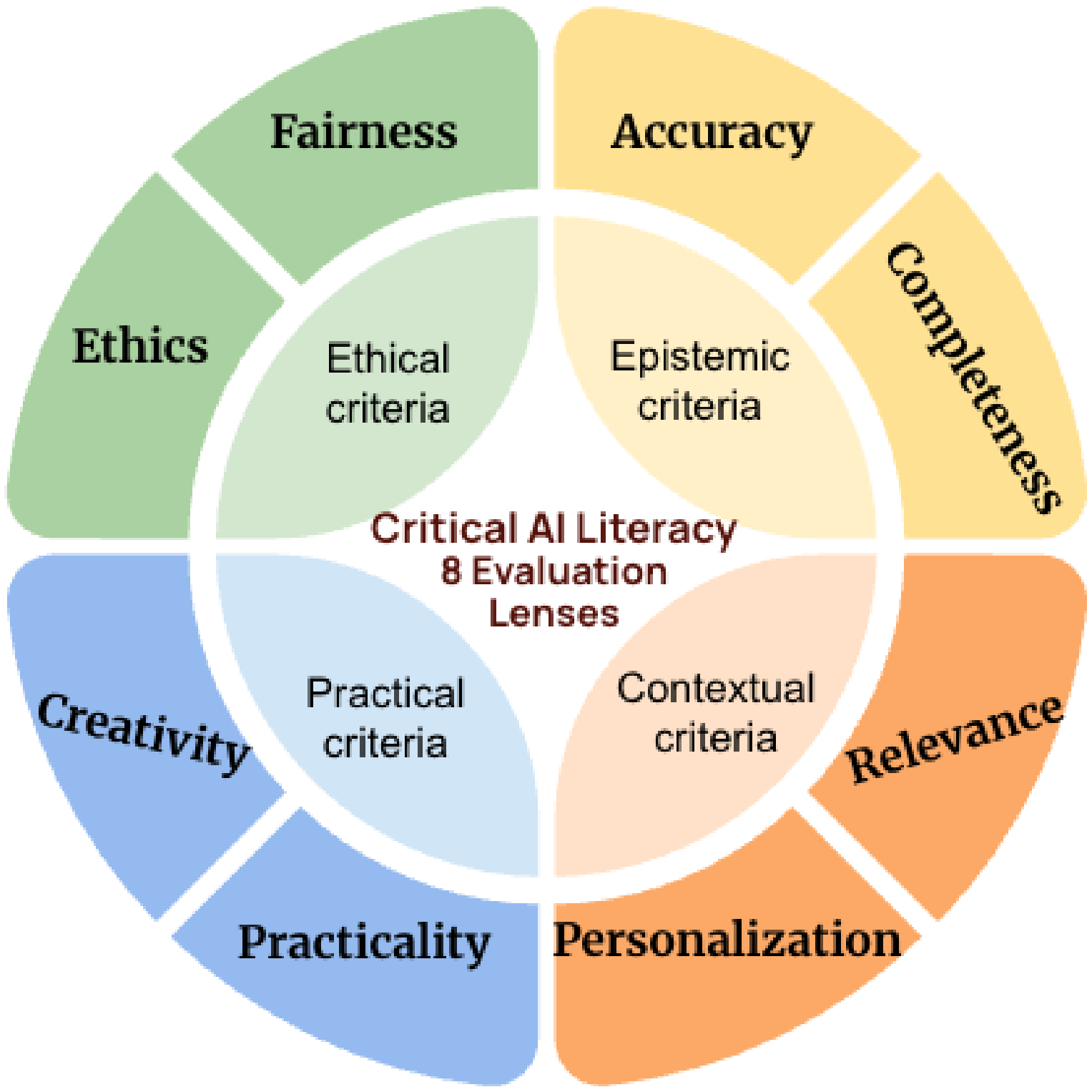

***4.3.1. Accuracy: Evaluating the Epistemic Reliability of AI Output***

Accuracy represented an epistemic criterion through which students assessed whether information provided by the chatbot was factually correct and trustworthy. Students explicitly indicated that AI-generated information should not be accepted without verification. For example, one student explained, "I was looking for accuracy to the topic. Honestly, I was really open to all suggestions with the intention to fact check and consider the logistics of certain implementations afterwards." This response illustrates that the student treated AI output as a provisional source of information rather than as an authoritative answer.

Other students evaluated responses based on whether they accurately explained the relevant concept. One student noted that "It explained the concept well," suggesting that perceived correctness and explanatory adequacy contributed to the student's judgment of the response. Collectively, these interactions demonstrate that students recognized the need to scrutinize the epistemic reliability of AI-generated content rather than accepting its claims at face value.

***4.3.2. Completeness: Evaluating the Sufficiency and Depth of AI Output***

Students also evaluated AI responses according to completeness, referring to whether the output provided sufficient detail, specificity, examples, or information to support their task. Rather than asking only whether an answer was correct, students considered whether it was *enough* for their particular purpose.

For example, students requested “more examples” when they perceived an AI response as insufficiently developed. Another student explicitly stated, “I’d like to check whether it would provide me with enough information.” In contrast, students sometimes described AI suggestions as “detailed” when the responses provided sufficient depth for their needs. This lens demonstrates that students' evaluation of AI extended beyond factual correctness to include judgments about the sufficiency and depth of information relative to the demands of the task.

#### *4.3.3. Task Alignment and Relevance: Evaluating Fit with the Assignment*

A third lens concerned task alignment and relevance. Students assessed whether AI-generated suggestions were connected to the purpose of their assignment, learning objectives, and task requirements. This criterion reflects an important distinction between a generally useful AI response and a response that is useful *for a specific academic task*.

Students explicitly articulated this criterion in statements such as, “I need relatability for my assignment” and “I wanted to check the relevance.” In some cases, students rejected AI suggestions because they did not satisfy the requirements of the task, as illustrated by the statement, “This isn't aligned with my assignment objective.” These responses indicate that students used the assignment itself as an evaluative reference point rather than allowing the chatbot to determine what constituted an appropriate response.

#### *4.3.4. Personalization: Evaluating Fit with the Learner, Audience, and Context*

Students further evaluated AI output according to its degree of personalization, particularly its fit with their intended audience, instructional context, and individual circumstances. AI-generated responses were not necessarily considered useful simply because they were well formulated; students also considered whether the suggestions reflected the specific people and contexts for whom their work was designed.

For example, one student stated, “I need it to match my audience,” while another explained that they evaluated “the AI suggestions based on my target audience.” Students also described modifying AI-generated suggestions when those suggestions did not sufficiently reflect their intended context. Thus, personalization functioned as a criterion for distinguishing between generic AI-generated recommendations and contextually meaningful solutions.

#### *4.3.5. Practicality: Evaluating Feasibility and Applicability*

Students also assessed whether AI-generated ideas could realistically be implemented. This practicality lens captured judgments about feasibility, appropriateness, available resources, and real-world constraints. Students questioned whether suggested activities or solutions could actually work within their intended instructional or organizational settings.

Examples included direct judgments such as “Not realistic,” “This wouldn't work,” and “The activity wasn't appropriate.” These responses demonstrate that students did not equate the plausibility of an AI-generated idea with its practical usefulness. Instead, they evaluated AI suggestions against the constraints of actual implementation, positioning their own contextual knowledge as an important source of judgment.

#### *4.3.6. Creativity: Evaluating Originality and Diversity of Ideas*

Another evaluation lens concerned creativity, through which students assessed whether AI-generated suggestions offered sufficiently original, varied, or distinctive possibilities. Students sometimes turned to the chatbot specifically to expand their idea space, but they also evaluated whether the resulting suggestions were sufficiently different from familiar or conventional approaches.

For example, students described wanting “something different” and characterized some AI-generated ideas as “too generic.” These responses suggest that students did not evaluate AI-generated ideas solely in terms of quantity or usefulness. They also considered whether AI contributed meaningful novelty to their own thinking. Creativity therefore served both as a criterion for evaluating the quality of generated ideas and for determining whether AI meaningfully expanded rather than merely reproduced conventional and general solutions.

#### *4.3.7. Bias and Fairness: Evaluating Inclusivity and Potential Disadvantage*

Students also evaluated AI-generated responses through a bias and fairness lens, considering whether the suggestions were neutral, inclusive, and equitable across different learners, audiences, or groups. This lens was distinct from general ethical evaluation because students specifically scrutinized the representation and distributional implications of AI-generated recommendations. Students directly questioned whether AI’s responses were neutral or biased, asking, “Is it neutral or biased?” Others explicitly sought ways to make their instructional designs more inclusive, asking, “What are your thoughts for the most inclusive experience?” Students also considered the potential consequences for particular groups, as reflected in the statement, “This could be a disadvantage… To ensure an inclusive experience…”

These interactions demonstrate that students did not simply evaluate whether an AI response was useful for themselves. They also considered who might be advantaged, disadvantaged, included, or excluded by implementing the AI-generated suggestion. Bias and fairness therefore constituted a distinct evaluative practice through which students examined the social and representational dimensions of AI output.

#### *4.3.8. Ethics and Responsibility: Evaluating Whether AI Use Is Appropriate and ethical*

In addition to evaluating bias and fairness, students assessed AI-generated responses through an ethics and responsibility lens. This criterion concerned whether a proposed use of AI or an AI-generated recommendation was ethically, legally, and responsibly appropriate to implement. Whereas the bias/fairness lens focused primarily on *how AI output represents or affects different groups*, the ethics/responsibility lens focused on the broader appropriateness and potential consequences of using AI in a particular situation. For example, one student explained, "I examined the responses whether it would cross an ethical line to investigate the subject on my own." Another student valued the inclusion of ethical and legal considerations in AI's response, describing such considerations as "critical when it comes to potential abuse" and suggesting stronger guardrails for legally questionable topics.

These responses indicate that students considered not only whether an AI-generated recommendation was accurate or useful, but also whether using or implementing that recommendation was responsible.

## 5. Discussion

### 5.1. Process of Human-AI Collaboration, Not a Single Act of AI Use

The first research question examined how students engage with generative AI during academic task completion. Our findings revealed seven stages of AI-supported academic work and five functional roles assumed by the chatbot. Importantly, these stages did not constitute a fixed sequence followed uniformly by all students. Rather, students selectively mobilized AI at different points in their work and repositioned the chatbot according to their immediate task needs. This finding extends existing conceptualizations of AI literacy that emphasize not only the ability to use AI but also the ability to interact and collaborate with AI systems (Long & Magerko, 2020; Ng et al., 2021). Conversational generative AI enables an iterative relationship in which learners can ask, inspect, redirect, revise, and evaluate AI-generated contributions. Our findings provide empirical evidence of this more dynamic form of human-AI engagement.

The five functional roles identified in this study further suggest that the contribution of AI to academic work is functionally fluid. Students did not consistently position the AI agent as a source of answers. Instead, they used it as a tutor when conceptual clarification was needed, an idea generator during ideation, a decision-support resource when considering possible directions, an editor during revision, and an evaluator when seeking feedback on their work.

More importantly, the findings shift attention from whether students use AI to how students position AI within their own cognitive work. This distinction is consequential given recent concerns that students may be capable of using GenAI for diverse academic activities while experiencing difficulties in critically evaluating AI-generated content (Lee et al., 2026; Tang et al., 2025; Velander et al., 2024). Our findings suggest that productive human-AI collaboration involves students maintaining an active role in determining what function AI should serve at different points in the task process. AI may generate possibilities, but the learner remains responsible for determining which possibilities are useful, appropriate, feasible, and worth incorporating.

Thus, the findings support an agentic view of human-AI collaboration in which AI contributes to the task while students retain responsibility for directing and evaluating that contribution. This extends existing AI literacy frameworks by illustrating how such agency may be enacted across the unfolding process of academic work.

### 5.2. Accepting AI Output Is Not Necessarily Uncritical, and Adapting It Is Not Necessarily More Critical

The second contribution concerns how critical engagement should be understood and measured. A particularly important finding was that students who adapted AI-generated suggestions did not demonstrate significantly higher information evaluation scores than students who accepted them. Although students who accepted AI suggestions reported significantly greater satisfaction with the AI responses, the difference in information evaluation was not statistically significant.

This finding challenges a potentially intuitive assumption that modifying AI output necessarily represents a more critical form of AI use than accepting it. Such an assumption treats behavioral outcomes as proxies for the quality of students' critical engagement. Our qualitative findings suggest that this interpretation is overly simplistic. Students who accepted AI suggestions sometimes provided explicit rationales for doing so. Our findings revealed that acceptance can itself be the result of an evaluative judgment. Conversely, students who adapted AI suggestions often retained substantial portions of the AI-generated content and modified it primarily to accommodate a particular audience, assignment, or learning context.

Accordingly, our findings suggest that critical AI literacy is better conceptualized as a property of the reasoning underlying AI-related decisions than as a property of the decision outcome itself. This distinction is important because existing work has increasingly called for learners to move beyond passive consumption of AI-generated content toward more critical and reflective engagement (Velander et al., 2024). Our findings provide a more specific empirical account of what such engagement can look like: not simply rejecting or modifying AI output, but articulating *why* a particular response is trustworthy, useful, appropriate, or in need of modification.

This finding also highlights studying critical AI literacy as an enacted practice rather than solely as a self-reported disposition. A learner may report valuing critical evaluation without demonstrating it in practice. Conversely, a learner may demonstrate critical evaluation through the reasoning embedded in seemingly ordinary decisions to accept or modify AI-generated suggestions. Examining these reasoning processes therefore provides a more sensitive lens for understanding students' evaluative agency in human-AI collaboration.

### 5.3. Critical AI Literacy Is Multidimensional and Situated

The third and central contribution of this study is the identification of eight evaluative lenses through which students assessed AI-generated content. This finding extends existing conceptualizations of critical AI literacy by providing a more fine-grained account of what learners actually attend to when evaluating AI-generated content. These lenses were organized into four broader dimensions. Prior scholarship has established evaluation as a central component of AI literacy (Ng et al., 2021; Veldhuis et al., 2025). However, these conceptualizations largely describe what learners should be able to do. Our findings move toward specifying the evaluative criteria through which such critical engagement is enacted during actual AI-supported work.

Most notably, students' evaluative practices extended well beyond factual accuracy. Accuracy and completeness represented an epistemic dimension through which students considered whether AI-generated information was correct, sufficiently detailed, and trustworthy. However, students also evaluated AI responses according to contextual criteria. They asked whether a response was relevant to their assignment, aligned with task objectives, and appropriate for their intended audience or instructional context. This finding is particularly important because it demonstrates that critical evaluation of AI is inherently situated. An AI-generated response cannot be evaluated adequately in isolation from the context in which it will be used. A response may be accurate but irrelevant, detailed but impractical, or creative but inappropriate for a particular audience. Students' use of task alignment and personalization as evaluative lenses therefore expands the notion of AI evaluation beyond epistemic verification toward contextual judgment.

The practical/design dimension further reinforces this point. Students evaluated whether AI-generated suggestions were feasible and whether they contributed sufficiently original or diverse ideas. These criteria suggest that students judged AI not only as an information source but also as a collaborator in design and problem solving. In this sense, critical engagement involves determining whether AI actually adds value to the learner's work rather than simply whether AI can produce a plausible response.

Finally, students evaluated AI-generated suggestions through ethical and social lenses, including ethics and fairness. Students explicitly questioned whether AI responses were neutral or biased, considered how suggestions might affect different groups, and examined whether proposed uses crossed ethical or legal boundaries. These findings resonate with the broader AI literacy literature emphasizing ethical considerations as a fundamental dimension of AI literacy (Ng et al., 2021) and with critical AI literacy scholarship calling attention to the sociotechnical assumptions and consequences embedded in AI systems (Velander et al., 2024; Veldhuis et al., 2025).

Overall, the eight lenses suggest that critical AI literacy is not simply the ability to identify whether an AI response is "right" or "wrong." Rather, it involves multidimensional, context-sensitive judgment about whether an AI-generated contribution is epistemically sound, sufficiently developed, contextually appropriate, practically viable, creatively valuable, socially equitable, and ethically responsible.

### 5.4. Toward an Operationalization of Critical AI Literacy

The eight evaluation lenses identified in this study contribute to an ongoing challenge in the AI literacy literature: translating the broad aspiration of critical AI literacy into an empirically observable construct. Existing scholarship has established the importance of critical evaluation, but the construct has often remained relatively broad (Velander et al., 2024; Veldhuis et al., 2025). Our findings suggest that critical AI literacy can be made more visible and observable by examining the criteria learners invoke when making judgments about AI-generated content. Rather than asking only whether students can evaluate AI, the proposed framework directs attention to *what they evaluate and on what grounds*.

Accuracy is an important component of critical evaluation, but it represents only one of several dimensions through which students judged AI output. Students also considered completeness, relevance, contextual fit, feasibility, creativity, fairness, and ethical responsibility. Thus, a student could critically evaluate an AI response even when the response was factually accurate, if the student recognized that it was inappropriate for the intended audience, infeasible to implement, insufficiently inclusive, or misaligned with the task objective. This multidimensionality also resonates with critical literacy traditions, which position learners as active interpreters who interrogate information in relation to broader social and contextual conditions (Freire, 1970; Luke, 2012). Applied to generative AI, critical literacy therefore involves not only interrogating *what AI says* but also asking whether the output should matter for this particular purpose, for these particular people, and under these particular conditions.

In this respect, the proposed framework moves critical AI literacy from an abstract competency toward an observable evaluative practice. The critical AI evaluation lenses provide dimensions along which students can exercise human judgment.

### 5.5. From Using AI to Judging AI

Our findings support a shift in how AI literacy may be conceptualized in educational settings. Much of the emerging discourse around generative AI in education has understandably focused on whether students can use AI effectively, including their ability to prompt and generate content. Such competencies remain important. However, our findings suggest that the more consequential question for human-AI collaboration may be what happens after AI produces an output. Students in this study did not simply receive AI-generated content. They evaluated it against multiple criteria and made decisions about whether and how it should inform their work. This finding complements prior calls for more critical questioning and evaluation in students' interactions with generative AI (Lee & Yu, 2026; Tang et al., 2025) while extending these calls by identifying concrete dimensions through which such questioning occurred. This conceptualization represents a movement from using AI as a functional tool, to critically evaluating AI-generated content, and ultimately to exercising agency as an active learner who determines whether, how, and to what extent AI should influence their own work (Figure 3). Although AI can generate content and offer recommendations, the learner retains the agency to evaluate, negotiate, and make decisions about how those contributions are incorporated into the final work.

**Figure 3**

*From Using AI to Exercising Agency: A Conceptualization of Critical AI Engagement*

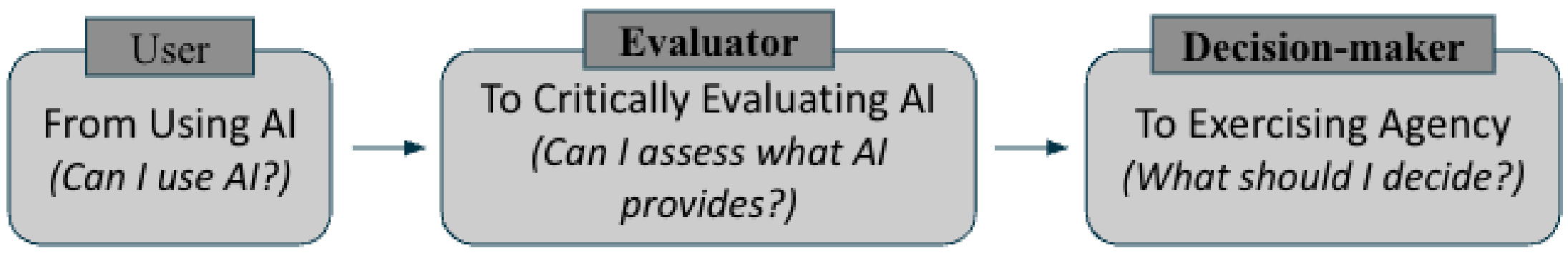

The eight evaluation lenses make visible the criteria through which learners exercised judgment and maintained ownership over decisions about AI-generated contributions. In doing so, the findings suggest that the goal of critical AI literacy is not to encourage learners to distrust AI indiscriminately, nor to encourage them to accept AI whenever it appears useful. Rather, it is to enable learners to make deliberate, context-sensitive, and reasoned judgments about AI-generated contributions.

This perspective also addresses a central tension in contemporary human-AI collaboration. AI can expand learners' access to ideas, explanations, feedback, and possible solutions, but increased access does not automatically produce better decisions. The educational value of AI therefore depends partly on learners' capacity to determine when an AI contribution is warranted, appropriate, and responsible. Critical AI literacy, from this perspective, is not an additional skill applied after AI use; it is a core mechanism through which human agency is maintained within AI-supported work.

## 5.6. Theoretical and Educational Implications

The findings have two primary implications for research on AI literacy. First, they contribute to the conceptualization of critical AI literacy by identifying a set of empirically grounded evaluative dimensions that can make the construct more specific and observable. The present findings specify the kinds of judgments learners may make while engaging with AI-generated content. Second, the findings contribute to scholarship on human-AI collaboration by highlighting evaluative agency as a central component of meaningful collaboration. Collaboration should not be understood simply as dividing cognitive labor between humans and AI. Rather, meaningful collaboration requires the human participant to retain the capacity to evaluate AI contributions and determine their role in the final product.

For educational practice, the framework provides a potential structure for helping students make their AI-related reasoning more explicit. Rather than providing generic instructions to "use AI critically" or "fact-check AI," educators could encourage students to interrogate AI-generated content through multiple dimensions: Is it accurate? Is it sufficiently complete? Is it relevant to my task? Does it fit my audience and context? Is it feasible? Does it add something creative or valuable? Is it fair and inclusive? Is its use ethical and responsible?

Importantly, these questions should not be understood as a rigid checklist through which every AI response must be evaluated identically. Rather, the eight lenses can function as a repertoire of evaluative perspectives that learners selectively mobilize according to the nature of the task and the potential consequences of using AI-generated content.

## 6. Conclusion

Overall, this study reframes critical AI literacy as a form of situated evaluative agency in human-AI collaboration. Our findings contribute to the emerging literature on critical AI literacy by offering an empirically grounded framework for understanding what critical evaluation looks like when learners actually interact with generative AI. The contribution is therefore not simply a list of criteria for evaluating AI outputs. Rather, the framework positions critical AI literacy as an enacted process in which learners draw on multiple evaluative lenses to determine whether AI-generated contributions are appropriate for their purposes and contexts. By making the evaluative dimensions of the judgment more visible, this study offers a foundation for future research examining how critical AI literacy develops, how it can be assessed, and how educational environments can support learners in maintaining agency within increasingly AI-mediated forms of learning and knowledge production.